\documentclass[superscriptaddress,preprint,jmp,amsmath,amssymb]{revtex4-1}

\usepackage{units}
\usepackage{graphicx}
\usepackage{dcolumn}
\usepackage{bm}
\usepackage[normalem]{ulem}
\usepackage{graphicx, datetime, MnSymbol, bbold}
\usepackage{amsmath} 
\usepackage{mathtools} 
\usepackage{amsfonts} 
\usepackage{color} 

\newcommand{\bra}[1]{\langle #1 \vert}
\newcommand{\ket}[1]{\vert #1 \rangle}

\begin{document}

\title{Distribution of high-dimensional entanglement via an intra-city free-space link}


\author{Fabian Steinlechner}
\email{These two authors contributed equally.}
\affiliation{Institute for Quantum Optics and Quantum Information - Vienna (IQOQI), Austrian Academy of Sciences, Vienna, Austria}

\author{Sebastian Ecker}
\email{These two authors contributed equally.}
\affiliation{Institute for Quantum Optics and Quantum Information - Vienna (IQOQI), Austrian Academy of Sciences, Vienna, Austria}

\author{Matthias Fink}
\affiliation{Institute for Quantum Optics and Quantum Information - Vienna (IQOQI), Austrian Academy of Sciences, Vienna, Austria}

\author{Bo Liu}
\affiliation{Institute for Quantum Optics and Quantum Information - Vienna (IQOQI), Austrian Academy of Sciences, Vienna, Austria}
\affiliation{School of Computer, NUDT, 410073 Changsha, China}

\author{Jessica Bavaresco}
\affiliation{Institute for Quantum Optics and Quantum Information - Vienna (IQOQI), Austrian Academy of Sciences, Vienna, Austria}

\author{Marcus Huber}
\affiliation{Institute for Quantum Optics and Quantum Information - Vienna (IQOQI), Austrian Academy of Sciences, Vienna, Austria}

\author{Thomas Scheidl}
\affiliation{Institute for Quantum Optics and Quantum Information - Vienna (IQOQI), Austrian Academy of Sciences, Vienna, Austria}

\author{Rupert Ursin}
\affiliation{Institute for Quantum Optics and Quantum Information - Vienna (IQOQI), Austrian Academy of Sciences, Vienna, Austria}
\affiliation{Vienna Center for Quantum Science and Technology (VCQ), Vienna, Austria}

\begin{abstract}
Quantum entanglement is a fundamental resource in quantum information processing and its distribution between distant parties is a key challenge in quantum communications. Increasing the dimensionality of entanglement has been shown to improve robustness and channel capacities in secure quantum communications. Here we report on the distribution of genuine high-dimensional entanglement via a 1.2-km-long free-space link across Vienna. We exploit hyperentanglement, that is, simultaneous entanglement in polarization and energy-time bases, to encode quantum information, and observe high-visibility interference for successive correlation measurements in each degree of freedom. These visibilities impose lower bounds on entanglement in each subspace individually and certify four-dimensional entanglement for the hyperentangled system. The high-fidelity transmission of high-dimensional entanglement under real-world atmospheric link conditions represents an important step towards long-distance quantum communications with more complex quantum systems and the implementation of advanced quantum experiments with satellite links.
\end{abstract}

\maketitle


\section*{Introduction}
The distribution of quantum entanglement between distant parties is one of the main technological challenges in the pursuit of a global-scale quantum Internet. Several proof-of-concept studies have already demonstrated high-fidelity transmission of photonic entanglement via terrestrial long-distance free-space links \cite{Ursin:2007,Fedrizzi:2009,Jin:2010}, and established the viability of employing optical satellite links for quantum communication on a global scale \cite{Scheidl:2013,Vallone:2014}, and beyond \cite{Rideout:2012}. However, until very recently, these experimental studies have been focused on bi-partite binary photonic systems, i.e. the simplest state space that can exhibit quantum entanglement. Specifically, polarization qubits have been the system of choice for free-space quantum communications for over a decade.

Encoding several qubits per transmitted photon increases channel capacity and yields significant benefits in the implementation of advanced quantum information processing protocols, such as improving resilience with respect to noise and eavesdropping in secure quantum communications  \cite{Bechmann:2000,Gisin:2002,Aolita:2007,Ali:2007,DAmbrosio:2012,Nunn:2013,Mower:2013,Graham:2015}. Hence, increasing the dimensionality of entangled quantum systems can be considered one of the next key technological steps towards the realization of more practical quantum information processing protocols in real world scenarios. Furthermore, from a fundamental physics point of view, the more diverse variations of non-classical correlations that are possible in a large state space also provide a platform for diverse quantum physics experiments \cite{Collins:2002,Dada:2011,Hendrych:2012,DAmbrosio:2013a}.

High-dimensional quantum information can be encoded in various photonic degrees of freedom (DOF), such as transverse orbital angular momentum (OAM) \cite{Langford:2004,Mirhosseini:2015,Malik:2016,Krenn:2017}, discrete photon arrival time bins \cite{Zhong:2015}, or continuous-variable energy-time modes \cite{Kwiat:1993,Xie:2015}. The transmission of classical OAM modes through turbulent atmosphere has been studied in several field trials \cite{Krenn:2014,Krenn:2016} and OAM multiplexing has already been used to achieve record channel capacity in free-space optical communications  \cite{Wang:2012}. While OAM entanglement has already been successfully demonstrated after atmospheric propagation \cite{Krenn:2015}, active wavefront correction will be required to fully exploit the potential of OAM encoding. The development of suitable adaptive optics systems is an immensely challenging field of ongoing research. Energy-time entanglement and its discrete analogue time-bin entanglement both offer alternatives for high-dimensional state encoding. Time-bin qubits \cite{Brendel:1999,Tittel:2000} have been routinely used in fiber-based quantum key distribution systems, which has culminated in the recent demonstration of quantum teleportation over long-distance fiber links \cite{Sun:2016,Valivarthi:2016} but have only recently been considered as a viable option for free-space quantum communications in presence of atmospheric turbulence \cite{Jin:2015,Vallone:2016}. 

The dimensionality of the state space can also be increased by simultaneously encoding quantum information in several DOF \cite{Simon:2014}. This has the significant advantage that single-photon two-qubit operations can be implemented deterministically between different DOF using only passive linear optics devices \cite{Fiorentino:2004a,Zhou:2015}. Furthermore, simultaneous entanglement in multiple degrees of freedom, known as hyperentanglement \cite{Kwiat:1997}, is readily engineered via the process of spontaneous parametric down-conversion (SPDC) in nonlinear crystals \cite{Barreiro:2005}. Hyperentanglement has been exploited in the realization of numerous advanced experiments, such as hyperentanglement-assisted Bell-state measurements \cite{Walborn:2003,Schuck:2006,Barbieri:2007,Jin:2010}, quantum teleportation of multiple DOF of a single photon \cite{Sheng:2010b,Wang:2015}, robust quantum communications with increased channel capacity \cite{Barreiro:2008}, and efficient entanglement purification schemes \cite{Simon:2002,Sheng:2010,Sheng:2010a}. However, experiments which exploit hyperentanglement have not yet ventured beyond the distance limitations of optical tables and protected laboratory environments.

In this article, we report on the distribution of energy-time and polarization hyperentangled photons via a 1.2-km-long intra-city free-space link. We observe high-visibility two-photon interference for successive correlation measurements in the polarization space and a 2-dimensional energy-time subspace and certify 4-dimensional entanglement in the combined system. Our assessment of energy-time entanglement is based on the observation of Franson interference in unbalanced polarization interferometers \cite{Franson:1989,Strekalov:1996}. This simple approach is highly suitable for the exploitation of such states in future quantum experiments with satellite links. 

\section*{Results}
\subsection{Experimental Setup}
\begin{figure*}
\centering
\includegraphics[width=\textwidth]{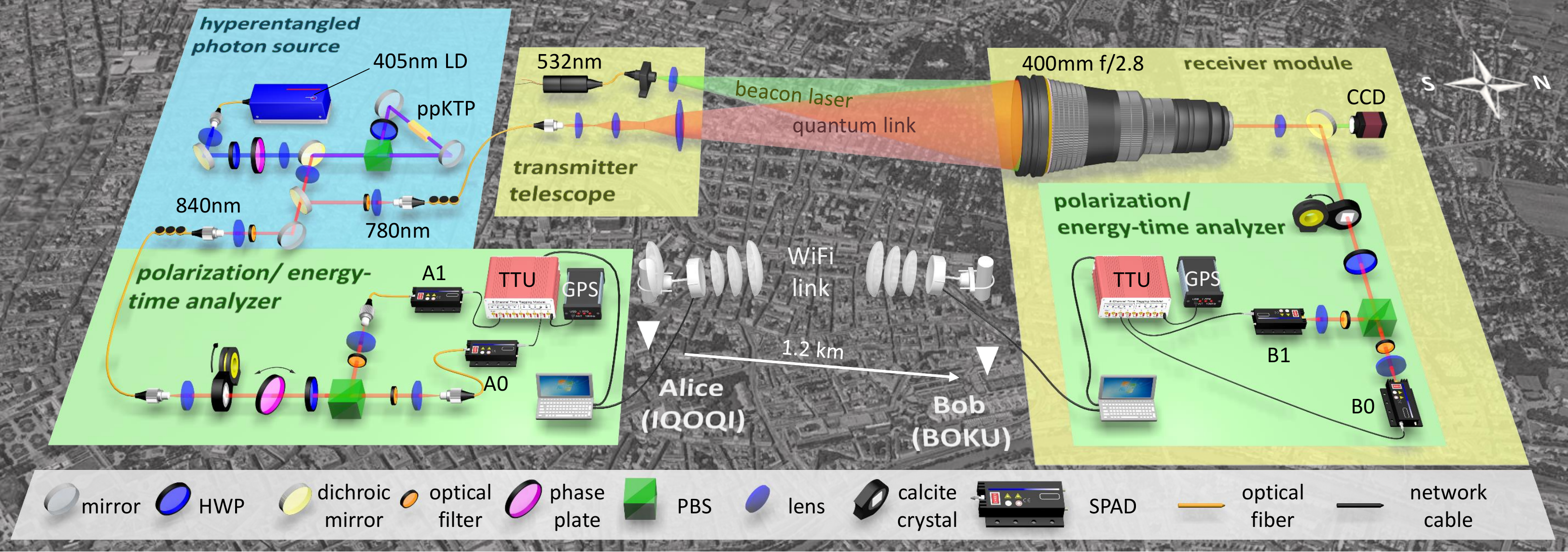} 
\caption{\label{expsetup} \textbf{Illustration of the high-dimensional entanglement distribution experiment.} A hyperentangled photon source was located in a laboratory at the Institute for Quantum Optics and Quantum Information Vienna (IQOQI). The source utilized spontaneous parametric down-conversion in a periodically poled KTiOPO$_4$ (ppKTP) crystal, which was placed at the center of a Sagnac interferometer and pumped with a continuous-wave 405-nm laser diode (LD). The polarization/energy-time hyperentangled photon pairs had center wavelengths of $\lambda_\text{B} \sim \unit[780]{nm}$ and $\lambda_\text{A} \sim \unit[840]{nm}$, respectively. Photon A was sent to Alice at IQOQI using a short fiber link, while  photon B was guided to a transmitter telescope on the roof of the institute and sent to Bob at the University of Natural Resources and Life Sciences (BOKU) via a 1.2-km-long free-space link. At Bob, the photons were collected using a large-aperture telephoto objective with a focal length of 400mm. The 532-nm beacon laser was separated from the hyperentangled photons using a dichroic mirror and focused onto a CCD image sensor in order to maintain link alignment and to monitor atmospheric turbulence. 
Alice's and Bob's analyzer modules allowed for measurements in the polarization or energy-time basis. The polarization was analyzed using a half-wave plate (HWP) and a polarizing beam splitter (PBS) with single-photon avalanche diodes (SPAD) in each output port. An additional phase shift could be introduced in Alice’s measurement module by tilting a birefringent crystal about its optical axis. In both analyzer modules, optional calcite crystals could be added before the PBS in order to introduce the polarization-dependent delay required for Franson interference measurements in the energy-time basis. Single-photon detection events were recorded with a GPS-disciplined time tagging unit (TTU) and stored on local hard drives for post-processing. Bob's measurement data was streamed to Alice via a classical WiFi link in order to identify photon pairs in real time. Map data \copyright 2017 Google.}
\end{figure*}

The experiment (depicted in Fig. \ref{expsetup}) was performed with an ultra-bright source of hyperentangled photons and a detection module (Alice) located at the Institute for Quantum Optics and Quantum Information (IQOQI) and a receiver station (Bob) at the University of Natural Resources and Life Sciences (BOKU) in Vienna. 

The source of hyperentangled photons (see methods section) was based on type-0 SPDC in a polarization Sagnac interferometer \cite{Kim:2006,Fedrizzi:2007} with a continous-wave 405-nm pump laser. It produced fiber-coupled polarization-entangled photon pairs with a two-photon coherence time ($t_\text{c} \lesssim \unit[1]{ps}$)  and center wavelengths $\lambda_\text{A} \sim \unit[840]{nm}$ and $\lambda_\text{B} \sim \unit[780]{nm}$, where subscripts A and B label the respective single-mode fiber for Alice and Bob. Since the emission time of a photon pair is uncertain within the significantly longer coherence time of the pump laser ($t_\text{p} \gtrsim \unit[100]{ns}$), photons A and B were entangled in energy-time\cite{Franson:1989}. In our proof of concept demonstration we focused on a two-dimensional subspace of the high-dimensional energy-time space (see methods). The total state space considered in our proof-of-concept experiment is thus a 4-dimensional hyperentangled state in polarization and energy-time: 

\begin{equation}
\ket{\Psi}_{\text{total}} =  \ket{\Phi}_{\text{pol}} \otimes \ket{\Phi}_{\text{e-t}} =  \frac{1}{2}\left( \ket{\text{H}}_\text{A} \ket{\text{H}}_\text{B}+\ket{\text{V}}_\text{A} \ket{\text{V}}_\text{B} \right)  \otimes   \left(\ket{t}_\text{A}\ket{t}_\text{B}  + \ket{t+\tau}_\text{A}\ket{t+\tau}_\text{B}     \right)
\end{equation}

where $\text{H}$ and $\text{V}$ represent horizontally and vertically polarized photon states whereas $t$ and $t+\tau$ denote photon-pair emission times with a delay $\tau$ with $t_\text{p}\gg\tau>t_\text{c}$. 

Photon A was guided to a local measurement module and photon B was guided to a transmitter telescope on the roof of the institute via a 15-m-long single-mode fiber. The photons emanating from the transmitter telescope were made to overlap with a 532-nm beacon laser for pointing, acquisition, and tracking (PAT) and sent to a receiver telescope at BOKU via a 1.2-km-long optical free-space link. The receiver telescope consisted of a telephoto objective (Nikkor $f=\unit[400]{mm}$ f/2.8) and an additional collimation lens. Note that the same type of objective is currently installed in the ISS Cupola module, and was recently proposed as a receiver in a quantum uplink scenario \cite{Scheidl:2013}. The beacon laser was transmitted through a dichroic mirror and focused onto a CCD image sensor while the collimated single-photon beam was guided to Bob's measurement module. 

The measurement modules for Alice and Bob each featured a polarization analyzer and an optional transfer setup that coupled the energy-time DOF to the polarization DOF (see also Supplementary Fig. 2). Alice's polarization analyzer consisted of a variable phase shifter, a half-wave plate, and a polarizing beam splitter (PBS) with multi-mode fiber-coupled single-photon avalanche diodes (SPAD) in each of its two output ports. A variable phase shift $\phi(\theta)$ could be introduced between the computational basis states $\ket{\text{H/V}}$ by tilting a birefringent YVO$_4$ crystal by an angle $\theta$ about its optical axis using a stepper motor. With the half-wave plate set to \unit[22.5]{\textdegree} this configuration corresponds to a polarization measurement in a superposition basis $+\phi$ / -$\phi$, where $\ket{\pm\phi}=\frac{1}{\sqrt{2}} \left( \ket{\text{H}}\pm e^{i\phi} \ket{\text{V}} \right)$. Bob's polarization analyzer module used a motorized half-wave plate and a PBS with a SPAD (active area of $\unit[180]{\mu m}$) in each of its two output ports. In order to reduce background counts from the city, long pass filters and interference filters were added and the optical system was engineered such that the detectors had a small field of view ($\unit[225]{\mu rad}$). Bob's analysis setup allowed for measurements in any linear polarization basis, in particular the basis +45\textdegree/-45\textdegree, where $\ket{\pm45^\circ}=\frac{1}{\sqrt{2}}\left(\ket{\text{H}}\pm\ket{\text{V}}\right)$. 

For the analysis of energy-time entanglement we employed a variant of the original Franson interferometer \cite{Franson:1989}, that uses polarization-dependent delays to map an energy-time subspace spanned by early $\ket{t}$ and late $\ket{t+\tau}$ pair emissions to the polarization state space \cite{Strekalov:1996}. This variant has the advantage that the polarization entanglement acts as a pair of synchronized switches, such that there is no need for detection post-selection \cite{Langford:PhD}. These unbalanced polarization interferometers at Alice and Bob were implemented with 3-mm-long calcite crystals, which could be inserted before the polarization analyzers. The calcite crystal introduced a birefringent delay of $\tau \sim \unit[2]{ps}$, which exceeded the coherence time of the SPDC photons but was significantly shorter than the coherence time of the pump laser. Note that the particular choice of delay restricts our considerations to a two-dimensional subspace of the intrinsically continuous-variable energy-time space. Hence, after introducing the polarization-dependent delay, polarization measurements in a superposition basis correspond to measurements of energy-time superpositions of the form $\ket{t}+e^{i{\phi}}\ket{t+\tau}$. A more detailed discussion is provided in the Supplementary Discussion. 

The arrival times of single-photon detection events at Alice and Bob were recorded relative to local 10 MHz GPS-disciplined clocks and stored on local hard drives for post-processing of two-photon detection events. Bob's measurement data was also streamed to Alice via a 5 GHz directional WiFi antenna where all combinations of two-photon detection events within a coincidence window of \unit[2]{ns} were monitored on-the-fly, while compensating for relative clock drifts (see Fig. \ref{fig:link-rate}) \cite{Ho:2009}.

\subsection{Link performance}
Directly at the source we detected a total coincidence rate of $R^{(2)}\sim \unit[84]{kcps}$ and singles rates of $R^{(1)}_\text{A} \sim \unit[400]{kcps}$, and $R^{(1)}_\text{B}\sim \unit[350]{kcps}$. Of the single photons sent via the free-space link we measured an average of $\unit[100]{kcps}$ in Bob's two detector channels, and an average rate of $\sim \unit[20]{kcps}$ two-photon detection events per second. For night-time operation the background counts were approximately $R^{(1)}_\text{B,1} \sim \unit[450-800]{cps}$ and $R^{(1)}_\text{B,0} \sim \unit[250-400]{cps}$ for Bob's two detector channels, whereby $\unit[200]{cps}$ and $\unit[50]{cps}$ were due to intrinsic detector dark counts.

\begin{figure}
\centering
\includegraphics[width=\columnwidth]{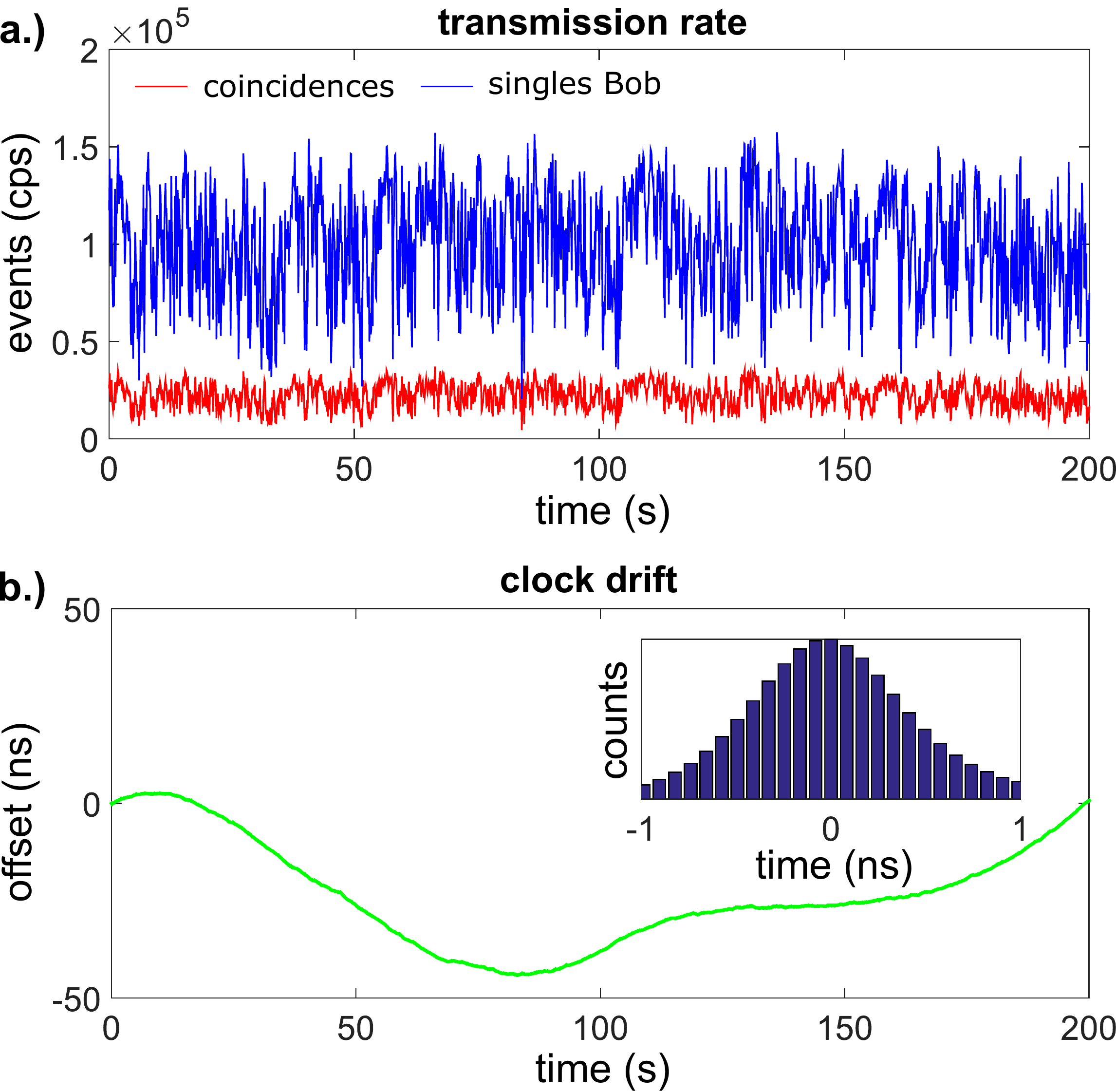}
\caption{\label{fig:link-rate} \textbf{Transmission rate and clock drift.} a.) Average single-photon (blue line) and two-photon (red line) detection rate (100ms integration time) after 1.2-km-long free-space transmission. The short-term signal fluctuated due to atmospheric turbulence, whereas the time-averaged rate of approximately 20 kcps remained almost constant over several hours. b.) Relative clock drift between Alice and Bob. The inset depicts the normalized histogram of two-photon detection events in \unit[80]{ps} time bins centered around the flight-time offset of $\sim \unit[3.94]{\mu s}$. All data acquired for nighttime operation on April 25$^{th}$-26$^{th}$ 2016.}
\end{figure}

Due to atmospheric turbulence the link transmission varied on the time-scale of ms (see Fig. \ref{fig:link-rate}). The time-averaged beam diameter  at the receiver was of the same order as the receiver aperture (14.5cm). We observed an average total link transmission of approximately 18\% including all optical losses from source to receiver, where approximately half of the transmission loss was due to absorption in optical components.

Besides being used for PAT, the CCD image sensor also monitored angle of arrival fluctuations caused by atmospheric turbulence \cite{Fried:65}. The full width half maximum of the angular variation at the telescope was estimated with a series of short exposure images and was in the order of $\sim \unit[25]{\mu rad}$ which corresponds to an atmospheric Fried parameter of approximately $\sim \unit[2]{cm}$ at $\unit[532]{nm}$. This is similar to that experienced in a free-space link over 144 km on the Canary islands \cite{Ursin:2007} and represents a worst case scenario in a satellite communication experiment through the atmosphere. Note that the angle of arrival fluctuations were significantly smaller than the detector's field of view; the fluctuation of detected count rates visible in Fig. \ref{fig:link-rate} stem from beam wander at the aperture of the receiver telescope. 

\subsection{Experimental visibilities}
\begin{figure}
\centering
\includegraphics[width=\columnwidth]{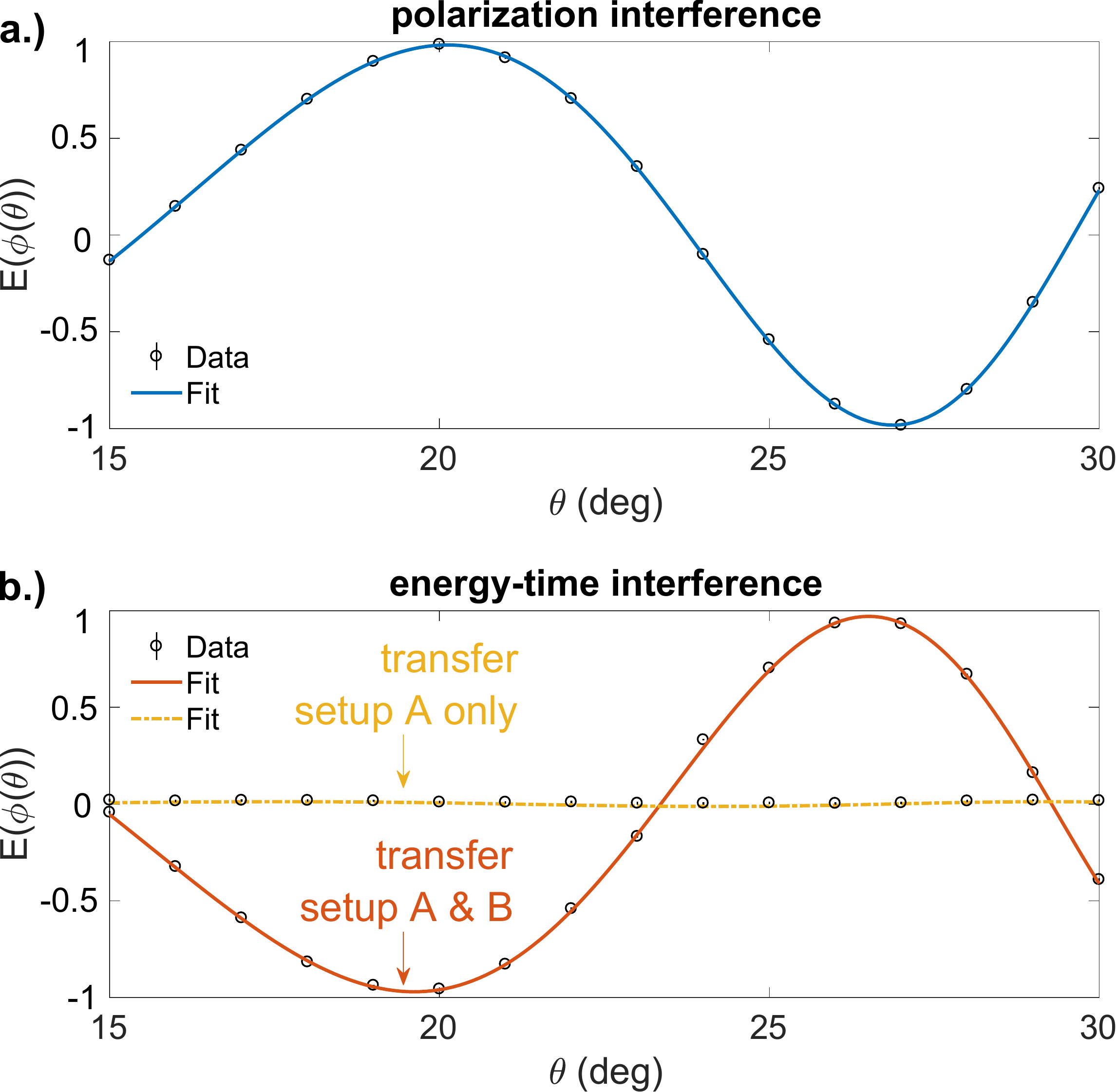} 

\caption{\label{fig:phase-scan} \textbf{Experimental characterization of hyperentanglement.} Two-photon correlation functions in the polarization basis (a.) and energy-time basis (b.) as a function of the variable phase shift introduced in Alice's measurement module. Each data point was evaluated from two-photon detection events accumulated over a 10 s integration time, without subtraction of accidental counts. The error bars which denote the 3-$\sigma$ standard deviation due to Poissonian count statistics are smaller than the data markers. The best fit functions (least-mean-square fit to the expected two-photon correlation in presence of experimental imperfections) exhibit visibilities $V^{\text{fit}}_{\text{pol}}=98.3\pm 0.5\%$ in the polarization basis (blue line) and $V^{\text{fit}}_{\text{e-t}}=96.8\pm 1\%$ in the energy-time basis (orange line). Almost no interference was observed when the energy-time to polarization transfer setup was introduced in Alice's detection module only (yellow line, $V^{\text{fit}}_{\text{e-t}}=1\pm 1\%$).}
\end{figure}

In order to verify the integrity of the atmospheric quantum communication channel for hyperentangled photons, we first assessed experimental two-photon polarization correlation:

\begin{equation} 
E=\frac{N_{1,1}+N_{0,0}-N_{1,0}-N_{0,1}}{\sum_{ij} N_{i,j}}
\end{equation}

where $N_{i,j}$ denotes the number of coincidence counts between Alice and Bob's SPAD detectors ($i,j \in \{1,0\}$). We define the experimental visibility $V$ in the superposition basis as the maximum correlation observed while scanning the phase of Alice's measurement basis  $+\phi$/$-\phi$ and keeping Bob's measurement setup in the linear +45\textdegree/-45\textdegree  polarization basis, i.e. $V=\text{max}_{\theta}(|E(\phi(\theta))|)$. The scan over the polarization correlations depicted in Fig. \ref{fig:phase-scan} exhibited a maximum value of $V_{\text{pol}}^{\phi}=98.5\pm 0.15\%$. The correlation in the linear $\text{H/V}$ measurement basis was $V^{\text{H/V}}_{\text{pol}}=99.33\pm0.015\%$. Note that the  $\text{H/V}$ visibility is limited almost exclusively due to accidental coincidences.

Similarly, with the transfer setup inserted in both measurement modules, we observed Franson interference with a visibility of $V_{\text{e-t}}^{\phi}=95.6\pm 0.3\%$ (Fig.\ \ref{fig:phase-scan}). In order to verify that the high visibility is due to two-photon energy-time entanglement, and not single-photon interference of photons A and B independently, we removed the transfer setup in Bob's detection module. In this case the measurement outcomes were completely uncorrelated, irrespective of $\phi(\theta)$, since the polarization-dependent delay exceeded the coherence time of the SPDC photons. This is indicated by the straight line in Fig. \ref{fig:phase-scan}. 

\subsection{Lower bounds on entanglement}
The experimental visibilities establish lower bounds of $0.978 \pm 0.0015$ and $0.912 \pm 0.006$ on the concurrence\cite{Huber:2011} in the polarization space and energy-time subspace, respectively (see methods). These values correspond to respective minimum values of $0.940\pm0.004$ and $0.776\pm0.014$ ebits of entanglement of formation. 

In the methods section we use these values to establish a lower bound for the Bell-state fidelity $\mathcal{F}(\hat{\rho}_{\text{pol,e-t}})$ of the hyperentangled state of the combined system (see also Supplementary Discussion). We achieve this by formulating this lower bound as a semidefinite programming problem, in which we minimize the $4$-dimensional concurrence and fidelity to a $4$-dimensional Bell state over all possible states in the combined Hilbert space that satisfy the experimentally observed subspace concurrences. 
We obtain lower bounds of $1.4671$ ebits of entanglement of formation and a Bell state fidelity of $0.9419$, thus certifying $4$-dimensional entanglement \cite{Fickler:2014}.

\section*{Discussion} 
We have distributed hyperentangled photon pairs via an intra-city free-space link under conditions of strong atmospheric turbulence. Despite the severe wave front distortions, we observed a high two-photon detection rate of $\sim \unit[20]{kcps}$ over a link distance of \unit[1.2]{km}. In a series of experiments we independently observed high-visibility two-photon interference in the 2-dimensional polarization state space and a 2-dimensional energy-time subspace. 
These visibilities are sufficient to certify  entanglement in both subspaces individually, and, for the first time, the coherent transmission of genuine high-dimensional quantum entanglement via a real-world free-space link. While the transmission of polarization-entangled photons has been studied in a number of previous field trials, our results demonstrate the feasibility of now also exploiting energy-time/polarization hyperentanglement in real-world link conditions with strong atmospheric turbulence.

Our analysis of interference in the energy-time DOF relies on an unbalanced polarization interferometer that coherently couples the polarization space with a 2-dimensional energy-time subspace. The current approach of mapping the time-bin entanglement to the polarization degree of freedom is of course intrinsically limited to accessing two-dimensional subspaces of the high dimensional energy time space. As recent experiments have clearly shown \cite{Tiranov:2016,Martin:2017}, the potential dimensionality of energy-time entanglement is orders of magnitudes larger. In fact, theoretically, it should only be limited by the achievable number of time-bins within the coherence time of the pump laser. The main challenge remains the implementation of superposition measurements, where a single calcite is inherently limited to two dimensions. Future setups for free-space experiments could use several delay lines, or a variable delay line \cite{Xie:2015} to greatly increase the dimensionality and with it the resistance to inevitable background noise.

Critically, the energy-time to polarization transfer setup can be understood as an implementation of a single-photon two-qubit operation \cite{Fiorentino:2004a}, which can be exploited in e.g. hyperentanglement-assisted Bell state measurements and efficient entanglement purification schemes \cite{Pan:2001,Simon:2002,Sheng:2010,Sheng:2010a}. In order to fully benefit from hyperentanglement in such applications, the delay between early and late photon arrival times will have to be directly resolved by the detectors. The main challenge therein lies in maintaining a constant phase relation between the long and short arms of the unbalanced interferometers for distorted input beams with a wide range of angles-of-incidence. However, such free-space compatible time-bin analyzers have recently been demonstrated in Refs.\cite{Jin:2015,Vallone:2016}, where the issue was ingeniously tackled via the implementation of a 4-$f$ imaging system in the long arm of the interferometer.

The coherent transmission of quantum information embedded in a genuine high-dimensional state space under real-world link conditions represents an important step towards long-distance quantum communications with more complex quantum systems and could play a key role in the implementation of advanced quantum information processing protocols in the future. A large quantum state space not only allows for larger information capacity in quantum communication links, as well as devising quantum communication schemes with more resilience against noise and improved security against eavesdroppers, but also allows for more diverse types of non-classical correlations which could prove vital in addressing technological challenges on the path towards global-scale quantum networks, as well as fundamental physics experiments.

Since polarization-entangled photon sources based on SPDC quite naturally exhibit energy-time entanglement when pumped with a continuous wave pump laser, the approach can readily be implemented with existing sources and proposals for satellite-link experiments with polarization-entangled photons without need for additional critical hardware \cite{Jennewein:2013,Merali:2012,Scheidl:2013,Tang:2016}. The additional possibility of analyzing energy-time entanglement could provide a platform for entirely new fundamental physics experiments with long-distance satellite links, such as the evaluation of models for gravity-induced wave function collapse \cite{Joshi:2017} or quantum information processing in a relativistic framework \cite{Bruschi:2014b}. High-dimensional energy-time entangled states can also be considered as a natural candidate for applications in quantum-enhanced clock synchronization protocols \cite{Giovannetti:2001}, and could allow for significant gains in performance by employing other quantum features, such as non-local cancellation of dispersion \cite{Giovannetti:2001a}. We also believe that our results will motivate both further theoretical research into energy-time entanglement experiments conceivable at relativistic scenarios with satellite links, as well as experimental research into the exploitation of hyperentanglement in long-distance quantum communications.

\section{methods}
\subsection{Hyperentangled photon source}
The hyperentangled photon source was based on type-0 spontaneous parametric down-conversion (SPDC) in a periodically poled KTiOPO$_4$ (ppKTP) crystal. The ppKTP crystal was bi-directionally pumped inside a polarization Sagnac interferometer \cite{Kim:2006,Fedrizzi:2007} and generated polarization-entangled photon pairs with center wavelengths $\lambda_{\text{A}} \sim \unit[840]{nm}$ and $\lambda_{\text{B}} \sim \unit[780]{nm}$. Photons A and B were separated using a dichroic mirror and coupled into optical single-mode fibers. For a pump power of 400 $\mu$W incident on the crystal, we detected a pair rate of of $R^{(2)}\sim \unit[84]{kcps}$ and singles rates of $R^{(1)}_\text{A} \sim \unit[400]{kcps}$, and $R^{(1)}_\text{B}\sim \unit[350]{kcps}$  directly after the source's single-mode fibers. This corresponds to a normalized detected pair rate of 200 kcps/mW and a detected spectral brightness of 100 kcps/mW/nm. Without correcting for background counts, losses, or detection inefficiency, we measure an average coincidence-to-singles ratio $R^{(2)}/\sqrt{R^{(1)}_\text{B} R^{(1)}_\text{A}}\sim 0.22$. 

The quasi-phase matching condition in the 20-mm-long ppKTP crystal \cite{Steinlechner:2014} resulted in a spectral bandwidth of $\Delta \lambda\sim \unit[2]{nm}$, which corresponds to a two-photon coherence time of $t_\text{c} \lesssim \unit[1]{ps}$. The emission time of a photon pair is uncertain within the significantly longer coherence time of the continuous-wave grating stabilized pump laser diode ($t_\text{p} \gtrsim \unit[100]{ns}$), such that the bi-photon state is in a superposition of possible pair-emission times (see also Supplementary Fig. 1), i.e. entangled in the energy-time DOF \cite{Franson:1989}. 

\subsection{Energy-time visibility measurement}
\begin{figure}
\centering
\includegraphics[width=0.5\textwidth]{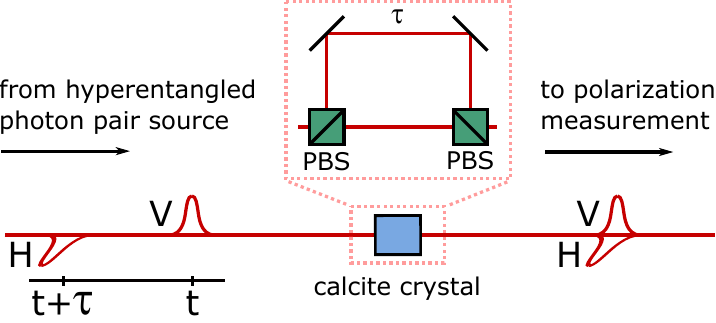} 
\caption{\label{fig:scheme-calcite}
\textbf{Energy-time to polarization transfer setup.} The calcite crystal acts as an unbalanced polarization interferometer which introduces a time shift of $\tau>t_\text{c}$ between vertically (V) and horizontally (H) polarized photons. After the transfer setup polarization measurements in a superposition basis allow to probe energy-time coherence (see also Supplementary Discussion).}
\end{figure}

We employed a variant of the original Franson scheme \cite{Strekalov:1996,Franson:1989} with unbalanced polarization interferometers to assess the coherence of the energy-time state. The polarization interferometers were implemented with birefringent calcite crystals, which introduced a polarization-dependent time shift $\tau$ (Fig. \ref{fig:scheme-calcite}). The particular choice of delay defines a 2-dimensional subspace (of the intrinsically continuous-variable energy-time space) spanned by the time-delayed basis states $\ket{t}$ and $\ket{t+\tau}$. Since this delay is significantly shorter than the timing resolution of the detectors, our experimental results can be understood as averages over a larger state space in the energy-time domain. The maximally-entangled Bell state in this subspace reads:

\begin{equation}
 \ket{\Phi}_{\text{e-t}} =\frac{1}{\sqrt{2}} \left(\ket{t}_\text{A}\ket{t}_\text{B}  + \ket{t+\tau}_\text{A}\ket{t+\tau}_\text{B}     \right)
\end{equation}

In the Supplementary Discussion (see also Ref. \cite{Langford:PhD}) we show how the transfer setup in combination with polarization entanglement is used to probe the experimental density matrix  ${{\rho}}'_{\text{e-t}}$ in the energy-time subspace. After introducing a polarization-dependent time shift for Alice and Bob, the visibility of polarization measurements in the superposition basis is determined by the off-diagonal coherence terms via: 

\begin{equation}
V^{\phi}_{\text{e-t}}\sim |{\bra{t,t} {{\rho}}'_{\text{e-t}} \ket{t+\tau, t+\tau}}|
\end{equation}
 
The total state space accessed in our experiment thus comprises the 2-dimensional polarization space and an effectively 2-dimensional energy-time subspace. The hyperentangled state of the total system can be expressed as a maximally-entangled state in four dimensions:

\begin{equation}
 \ket{\Psi}_{\text{total}} =  \frac{1}{2}\left( \ket{0}_\text{A}\ket{0}_\text{B} +  \ket{1}_\text{A}\ket{1}_\text{B} + \ket{2}_\text{A}\ket{2}_\text{B} + \ket{3}_\text{A}\ket{3}_\text{B}\right) 
\end{equation}

with basis vectors $\ket{0}=\ket{\text{H},t},\ket{1}=\ket{\text{H},t+\tau},\ket{2}=\ket{\text{V},t},\ket{3}=\ket{\text{V},t+\tau}$. For more details refer to the the Supplementary Discussion.

\subsection{Certification of entanglement}
In Ref. \cite{Huber:2011} easily computable lower bounds for the concurrence of mixed states that have an experimental implementation were derived:

\begin{equation}\label{methods2-1}
\mathcal{C}(\rho)\geq 2\times\text{Re}\left(\bra{00}\rho\ket{11}\right)-\left(\bra{01}\rho\ket{01}+\bra{10}\rho\ket{10}\right)
\end{equation}

where $\rho$ is the density matrix in the 2-dimensional subspace. In the Supplementary Discussion we show the concurrence can be related to the experimental polarization space and energy-time visibilities via:

\begin{equation}\label{methods2-2}
\begin{split}
\mathcal{C}(\rho_\text{pol}) & \geq V_{\text{pol}}^{\phi} + V_{\text{pol}}^{\text{H/V}}-1\\
\mathcal{C}(\rho_\text{e-t}) & \geq2\times V_{\text{e-t}}^{\phi} -1
\end{split}
\end{equation}
 
Note that the bound on the energy-time concurrence involves the additional assumption that there is no phase relationship between accidental coincidence that occur in time bins separated by more than the coherence time. We believe that, while this assumption precludes a certification of entanglement that meets the requirements for quantum cryptography, it is completely justified for our proof of concept experiment. This also agrees with our experimental observation that scanning the phase of the entangled state in the source had no effect on the single-photon coherence. 

With the experimentally obtained lower bounds for $\mathcal{C}(\rho_\text{pol})$ and $\mathcal{C}(\rho_\text{e-t})$ at hand, we calculate a lower bound for the concurrence of the global state $\mathcal{C}({\rho}_{\text{pol,e-t}})$ by solving the following convex optimization problem: a minimization of the function that defines a lower bound for the concurrence, over all states $\rho$ acting on a $4$-dimensional Hilbert space such that the concurrence of the reduced states in $2$-dimensional subspaces satisfy the constraints of being lower bounded by the values $\mathcal{C}(\rho_\text{pol})$ and $\mathcal{C}(\rho_\text{e-t})$. As demonstrated in the Supplementary Discussion, this convex optimization problem has a semidefinite programming (SDP) characterization and satisfies the condition of strong duality. Hence, the obtained lower bound of $\mathcal{C}({\rho}_{\text{pol,e-t}})\geq1.1299$ has an analytical character.

Another useful measure of entanglement is the entanglement of formation $E_\text{oF}(\rho)$, which represents the minimal number of maximally entangled bits (ebits) required to produce $\rho$ via an arbitrary local operations and classical communication (LOCC) procedure.
It can be shown \cite{Huber:2013} that the entanglement of formation is lower bounded by the concurrence according to:
\begin{equation}
E_\text{oF}(\rho)\geq-\log\left(1-\frac{\mathcal{C}(\rho)^2}{2}\right).
\end{equation} 
Hence, from the lower bound for the concurrence $\mathcal{C}({\rho}_{\text{pol,e-t}})$ it is possible to calculate a lower bound of $E_\text{oF}({\rho}_{\text{pol,e-t}})\geq1.4671$ for the entanglement of formation, which is sufficient to certify $3$-dimensional bipartite entanglement \cite{Huber:2013}.

By adapting the objective function of our SDP from the concurrence to the fidelity to the maximally entangled $4$-dimensional state, it is possible to lower bound the latter quantity by performing a minimization over the same variable and same constraints. As shown in the Supplementary Discussion, this second SDP also satisfies strong duality and provides the analytical bound of $\mathcal{F}({\rho}_{\text{pol,e-t}})\geq0.9419$, which certifies $4$-dimensional bipartite entanglement \cite{Fickler:2014}.

\section*{Acknowledgements}
We thank Johannes Handsteiner, Dominik Rauch, and S\"{o}ren Wengerowsky for their support in setting up the experiment. We also thank Mario Krenn and Sven Ramelow for helpful conversations and comments on the initial draft of the manuscript. We also thank the  Bundesimmobiliengesellschaft (BIG) for providing the room for our receiving station.
 
Financial support from the Austrian Research Promotion Agency (FFG) - Agentur f\"{u}r Luft- und Raumfahrt (FFG-ALR contract 844360), the European Space Agency (ESA contract 4000112591/14/NL/US), the Austrian Science Fund (FWF) through (P24621-N27) and the START project (Y879-N27), as well as the Austrian Academy of Sciences is gratefully acknowledged.
 
 \section*{Competing Interests} The authors declare that they have no competing financial interests.
 
\section*{Author Contributions} FS conceived the experiment. SE designed and developed the entangled photon source and local detection module under the supervision of FS. MF designed and developed the free-space link and receiver optics with help from TS. FS, SE, and MF performed the experiment under the guidance of RU. MF and BL designed the coincidence tracking software and processed detection events. FS, MF, and, SE analyzed the experimental data. FS, MF, TS, SE, and RU discussed and evaluated the experimental results. MH and JB established the bounds on high-dimensional entanglement and provided theory support. FS wrote a first draft and all authors contributed to the final version of the manuscript.  

 \section*{Author Contributions} Correspondence and requests for materials should be addressed to FS (fabian.steinlechner@oeaw.ac.at) and RU (rupert.ursin@oeaw.ac.at).

\end{document}


\title{Supplementary Information for \\Distribution of high-dimensional entanglement via an intra-city free-space link}


\author{Fabian Steinlechner}
\affiliation{Institute for Quantum Optics and Quantum Information - Vienna (IQOQI), Austrian Academy of Sciences, Vienna, Austria}

\author{Sebastian Ecker}
\affiliation{Institute for Quantum Optics and Quantum Information - Vienna (IQOQI), Austrian Academy of Sciences, Vienna, Austria}

\author{Matthias Fink}
\affiliation{Institute for Quantum Optics and Quantum Information - Vienna (IQOQI), Austrian Academy of Sciences, Vienna, Austria}

\author{Bo Liu}
\affiliation{Institute for Quantum Optics and Quantum Information - Vienna (IQOQI), Austrian Academy of Sciences, Vienna, Austria}
\affiliation{School of Computer, NUDT, 410073 Changsha, China}

\author{Jessica Bavaresco}
\affiliation{Institute for Quantum Optics and Quantum Information - Vienna (IQOQI), Austrian Academy of Sciences, Vienna, Austria}

\author{Marcus Huber}
\affiliation{Institute for Quantum Optics and Quantum Information - Vienna (IQOQI), Austrian Academy of Sciences, Vienna, Austria}

\author{Thomas Scheidl}
\affiliation{Institute for Quantum Optics and Quantum Information - Vienna (IQOQI), Austrian Academy of Sciences, Vienna, Austria}

\author{Rupert Ursin}
\affiliation{Institute for Quantum Optics and Quantum Information - Vienna (IQOQI), Austrian Academy of Sciences, Vienna, Austria}
\affiliation{Vienna Center for Quantum Science and Technology (VCQ), Vienna, Austria}
\maketitle

In the main manuscript we report on the distribution of polarization/energy-time hyperentangled photons via a free-space link. In order to assess the integrity of the atmospheric quantum communication channel we performed successive correlation measurements in superposition bases. We observed two-photon polarization interference visibility of $V_{\text{pol}}\sim 98.5\%$ and $V_{\text{e-t}}\sim95.6\%$ for Franson interference, respectively. 

While high-visibility interference in both degrees of freedom is a clear sign for the presence of hyperentanglement, these measurements do not yet quantify entanglement in the two subspaces, nor do they quantify the dimensionality of entanglement in the combined state space. Ideally, Alice and Bob would perform a complete tomography of the hyperentangled state. However, as we will show in the following, a complete state tomography is not required to obtain lower bounds on several quantitative measures of entanglement. Moreover, these lower bounds can be derived from independent visibility measurements in the energy-time and polarization subspaces, as shown in the final section of this theory supplement.

The supplementary discussion is structured as follows: First, we review the experimental setup used to measure two-photon visibilities in the energy-time and polarization degrees of freedom. Then, we show how the independently observed visibilities can be used to determine lower bounds on the concurrence and entanglement of formation in the  polarization and energy-time subspaces. We conclude with a discussion of the lower bound these measurements impose on the dimensionality of entanglement in the combined state space of the entire system.

\section*{Revision of Experimental Setup}
In this section we provide additional information on the experimental setup and methods used to quantify entanglement. We start by reviewing the nature of the hyperentangled target state, experimental density matrices, and visibility measurements in a notation that is more in line with the standard quantum information formalism. 

\subsection{Hyperentangled target state}
\begin{figure}[h]
\centering
\includegraphics[width=12cm]{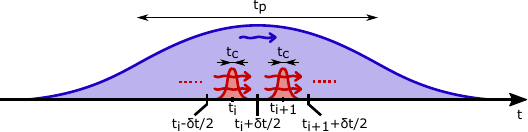} 
\caption{\label{fig:coherence} Illustration of relevant coherence times in energy-time space. The pump photon is indicated by a blue envelope with coherence time $t_\text{p}$, while two possible  SPDC photon pair emission times are indicated by red envelopes with coherence times $t_\text{c} \ll t_\text{p}$. Note, that the coherence times are not drawn to scale. For our analysis of the energy-time entangled state, the emission time of the photon pairs are grouped into non-overlapping time bins of length $\delta t$.}
\end{figure}

First, let us consider the ideal polarization/energy-time hyperentangled state: A strong cw pump laser with a coherence time $t_\text{p}$ pumps a single nonlinear crystal inside a polarization Sagnac interferometer and produces polarization-entangled photon pairs with a coherence time $t_\text{c}\lesssim1$ps. The coherence time of the pump laser is significantly longer than the coherence time of the signal and idler photons $t_\text{c}\ll t_\text{p}$, resulting in a maximally hyper-entangled state: 

\begin{equation}
 \ket{\Psi}_{\text{total}} =  \ket{\Phi}_{\text{pol}} \otimes \ket{\Phi}_{\text{e-t}} =  \frac{1}{\sqrt{2}}\left( \ket{\text{H}}_\text{A} \ket{\text{H}}_\text{B}+\ket{\text{V}}_\text{A} \ket{\text{V}}_\text{B} \right)  \otimes  \int \ket{\tau}_\text{A}\ket{\tau}_\text{B} \text{d}\tau
\end{equation}

where $\text{H}$ and $\text{V}$ represent horizontally and vertically polarized photon states, $\tau$ denotes the photon emission time, and the subscripts A and B label the respective single-mode fiber for Alice and Bob. For the sake of brevity we have assumed a cw pump laser with infinite coherence time and perfectly correlated photon pair emissions\footnote{For a more rigorous discussion of the energy-time state in SPDC, see e.g. Ref. \cite{Shih:2003}}. In the following discussion it will be sufficient to group pair emissions within the range $\tau=t_i-\delta t/2 \rightarrow \tau=t_i+\delta t/2$ into time bins $\ket{t_i}$ (see Fig.\ref{fig:coherence}). The size of these time bins is chosen to be larger than the coherence time of the SPDC photons, but smaller than the coherence time of the pump laser ($t_\text{c} \leq \delta t \ll t_\text{p} $). Hence, the energy-time state can be considered as a coherent superposition of $N \sim t_\text{p} / \delta t$ orthogonal time bins, thus constituting an $N$-dimensionally entangled state in the energy-time domain:

\begin{equation}\label{eq:energy-time-pure-state}
 \ket{\Phi}_{\text{e-t}} =\frac{1}{\sqrt{N}} \sum_{i=1}^{N} \ket{t_i}_\text{A}\ket{t_i}_\text{B}
\end{equation}

For large $N$ it is generally not experimentally feasible to exploit the entire state space and in our proof of concept demonstration we focused on a two-dimensional subspace spanned by time-delayed states  $\ket{t}$ and $\ket{t+\tau}$:
\begin{equation}
 \ket{\Phi}_{\text{e-t}} =\frac{1}{\sqrt{2}} \left(\ket{t}_\text{A}\ket{t}_\text{B}  + \ket{t+\tau}_\text{A}\ket{t+\tau}_\text{B}     \right)
\end{equation}

 The total state space accessed in our experiment thus comprises the 2-dimensional polarization space and an effectively 2-dimensional energy-time subspace.  More details on the definition of this state space will be provided in the following.  Omitting phase dependencies the hyperentangled state of the total system can be expressed as a maximally-entangled Bell state in four dimensions:

\begin{equation}
 \ket{\Psi}_{\text{total}} =  \frac{1}{2}\left( \ket{0}_\text{A}\ket{0}_\text{B} +  \ket{1}_\text{A}\ket{1}_\text{B} + \ket{2}_\text{A}\ket{2}_\text{B} + \ket{3}_\text{A}\ket{3}_\text{B}\right) 
\end{equation}

where $\ket{0}=\ket{\text{H},t} \ , \ \ket{1}=\ket{\text{H},t+\tau} \ , \ \ket{2}=\ket{\text{V},t} \ , \ \ket{3}=\ket{\text{V},t+\tau}$.

\section*{Polarization interference visibility and characteristic density matrix elements}
\begin{figure}
\centering
\includegraphics[width=14cm]{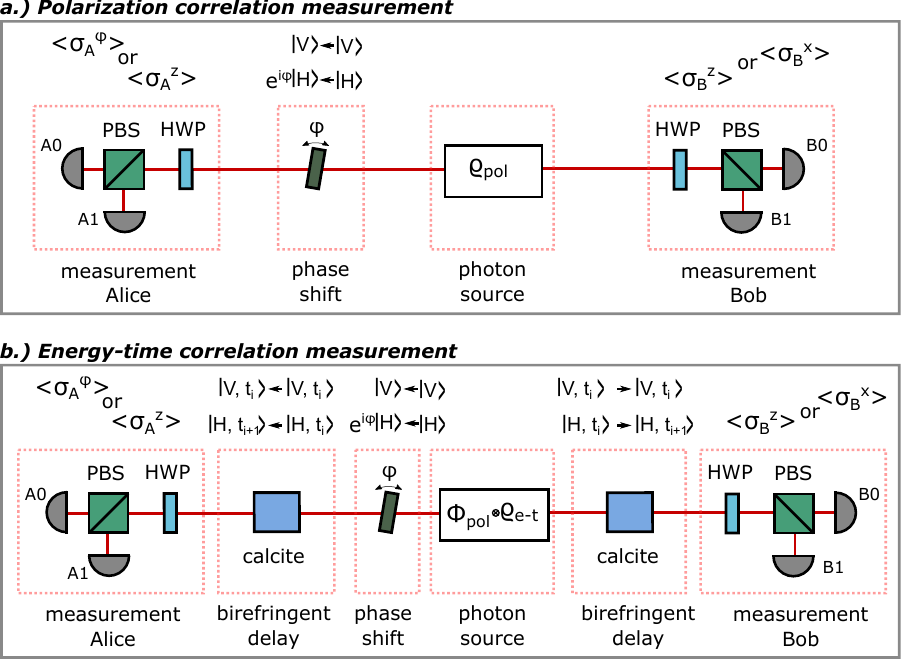} 
\caption{\label{fig:scheme-v2} Experimental setup for polarization and energy-time correlation measurements. Hyperentangled  photons are distributed to Alice and Bob. Alice applies an additional phase shift $\phi$ by tilting a birefringent crystal. Alice and Bob evaluate the visibility of the two-photon correlation functions in the computational basis $V^{\text{H}/\text{V}}_{\text{pol}}$=$\langle {\sigma}_\text{A}^\text{z} \otimes {\sigma}_\text{B}^\text{z} \rangle$ and the superposition basis $V^{\phi}_{\text{pol}}$=$\text{max}_\phi[\langle{\sigma}_\text{A}^\phi \otimes {\sigma}_\text{B}^\text{x}\rangle]$.
a.) Polarization correlation measurements b.) Energy-time correlation measurement: In order to assess the energy-time coherence an additional calcite crystal is added in Alice's and Bob's measurement setup. The calcite crystal introduces a polarization-dependent delay that couples the polarization and energy-time DOF, i.e. maps coherence in the energy-time state to coherence in the polarization DOF.}
\end{figure}

Our experimental implementation (see Fig.\ \ref{fig:scheme-v2}a.) allowed for measurements of the following observables:

\renewcommand{\arraystretch}{1.3}

\begin{tabular}{|r |c| c|}
\hline
  \text{} & \text{Observable} & \text{Eigenstates}  \\
  \hline

  \text{computational basis Alice} & ${\sigma}^\text{z}_\text{A} $&$ \ket{\text{H}/\text{V}}$ \\
  \text{computational basis Bob} & ${\sigma}^\text{z}_\text{B}$  & $\ket{\text{H}/\text{V}}$ \\
  \text{superposition basis Alice} &  ${\sigma}^\phi_\text{A}=\cos(\phi){\sigma}^\text{x}_\text{A}+\sin(\phi){\sigma}^\text{y}_\text{A}$ & $\ket{\pm\phi}=\frac{1}{\sqrt{2}} \left( \ket{\text{H}}\pm e^{i\phi} \ket{\text{V}} \right)$   \\
  \text{superposition basis Bob} & ${\sigma}^\text{x}_\text{B}$ &$ \ket{\pm45^\circ}=\frac{1}{\sqrt{2}}\left(\ket{\text{H}}\pm\ket{\text{V}}\right)$  \\
  \hline
\end{tabular}

where ${\sigma}^i_{\text{A}/\text{B}}$ denote the Pauli operators for Alice and Bob, respectively\footnote{Note that the linear $\pm 45^{\circ}$ basis is included as the special case ${\sigma}^{\phi=0}_\text{A}={\sigma}^\text{x}_\text{A}$.}. A full tomographic reconstruction of the polarization density matrix:

\begin{equation}
{\rho}_{\text{pol}} =
\bordermatrix{
  & \text{H}_\text{A}\text{H}_\text{B}	& \text{H}_\text{A}\text{V}_\text{B}   & \text{V}_\text{A}\text{H}_\text{B}  &  \text{V}_\text{A}\text{V}_\text{B}    \cr
 \text{H}_\text{A}\text{H}_\text{B} & \rho_{0000} & \rho_{0001} & \rho_{0010} & \rho_{0011} \cr
\text{H}_\text{A}\text{V}_\text{B} & \ddots &  \rho_{0101} &  \rho_{0110} & \rho_{0111} \cr
\text{V}_\text{A}\text{H}_\text{B} & \ddots & \ddots & \rho_{1010} & \rho_{1011} \cr
\text{V}_\text{A}\text{V}_\text{B} & c.c. & \ddots & \ddots & \rho_{1111} \cr}
\end{equation}

is not possible from these measurements. It is, however, possible to infer lower bounds on the absolute value of various density matrix elements. In order to see this, we now consider how the measured visibilities relate to elements of the density matrix. For the visibility of joint measurements in the computational basis we have:

\begin{equation}\label{rho_pol_Visibility_HV}
V^{\text{H}/\text{V}}_{\text{pol}}=\langle {\sigma}^\text{z}_\text{A} \otimes {\sigma}^\text{z}_\text{B} \rangle =\rho_{0000} - \rho_{0101} - \rho_{1010} + \rho_{1111}
\end{equation}

Similarly, for the visibility in the coherent superposition basis, which is mutually unbiased to the computational basis, we have:

\begin{equation}\label{rho_pol_Visibility_phi}
 V_{\text{pol}}^{\phi}= \text{max}_{\phi}[\langle {\sigma}^{\phi}_\text{A} \otimes {\sigma}^\text{x}_\text{B} \rangle]= 2\, \text{max}_\phi [\operatorname{Re}(\rho_{0011}e^{i\phi} +  \rho_{0110}e^{i\phi}) ]
\end{equation}
 
\subsection{Energy-time interference visibility and characteristic density matrix elements}
Next, let us discuss the correlation measurements that were performed in the energy-time subspace and relate the Franson interference visibility to the magnitude of certain matrix elements of the energy-time density matrix. For the sake of brevity let us assume that the state of the total system ${\rho}_{\text{pol,e-t}}$ is a product of energy-time and polarization states:  

\begin{equation}
{\rho}_{\text{pol,e-t}}=\ketbra{\Phi_{\text{pol}}^+}{\Phi^+_{\text{pol}}} \otimes {{\rho}}_{\text{e-t}}
\end{equation}

This assumption will be justified a posteriori when we demonstrate that the experimental polarization state has a high degree of overlap with the maximally entangled Bell state, as this also implies that the polarization degree of freedom (DOF) cannot exhibit significant correlations with the energy-time space.

As discussed in the main manuscript, we employed a variant of the original Franson scheme \cite{Strekalov:1996,Franson:1989} with unbalanced polarization interferometers to assess energy-time interference in a 2-dimensional subspace. These unbalanced polarization interferometers at Alice and Bob were implemented with 3-mm-long calcite crystals, which could be inserted before the polarization analyzers (see Fig.\ \ref{fig:scheme-v2}b.). The calcite crystals introduce a polarization-dependent time shift ${\tau}$. In the following we omit additional phase shifts due to the propagation through the crystal, as they can be absorbed into the phase plate in Alice's detection module. The operator describing the transformation in the polarization and energy-time space can thus be written as:

\begin{equation}
{\mathcal{T}}=\ketbra{\text{H}}{\text{H}} \otimes {\hat{\tau}} + \ketbra{\text{V}}{\text{V}} \otimes {1}
\end{equation}

 where  ${1}$ is the identity operator and $\hat{\tau}$ transforms the energy-time basis states (Eq. \ref{eq:energy-time-pure-state}) as:

\begin{equation}
\hat{\tau} \ket{t_i} = \ket{t_{i+1}}
\end{equation}

 Note that the particular choice of delay $\tau$ introduced in the unbalanced polarization interferometer restricts our considerations to a two-dimensional subspace of the intrinsically continuous-variable energy-time space, which could be fully exploited, e.g., in experimental implementations with a variable delay line \cite{Xie:2015,Martin:2017}. Since all the measurement results obtained in our experiment can be related to this 2-dimensional subspace, we define the effective energy-time density matrix as: 

 \begin{equation}
{{\rho}}_{\text{e-t}}'
= \begin{pmatrix}
\sum_{i} \bra{t_i t_i}\rho_{\text{e-t}}\ket{t_i t_i}  & \sum_{i}\bra{t_i t_i} \rho_{\text{e-t}}\ket{t_i t_{i+1}} & \sum_{i}\bra{t_i t_i} \rho_{\text{e-t}}\ket{t_{i+1} t_{i}} & \sum_{i}\bra{t_i t_i} \rho_{\text{e-t}}\ket{t_{i+1} t_{i+1}} \cr
  \ddots  & \sum_{i}\bra{t_i t_{i+1}} \rho_{\text{e-t}}\ket{t_{i} t_{i+1}} &  \sum_{i}\bra{t_i t_{i+1}} \rho_{\text{e-t}}\ket{t_{i+1} t_{i}} & \sum_{i}\bra{t_i t_{i+1}} \rho_{\text{e-t}}\ket{t_{i+1} t_{i+1}} \cr
\ddots & \ddots & \sum_{i}\bra{t_{i+1} t_{i}} \rho_{\text{e-t}}\ket{t_{i+1} t_{i}} & \sum_{i}\bra{t_{i+1} t_{i}} \rho_{\text{e-t}}\ket{t_{i+1} t_{i+1}} \cr
c.c. &  \ddots & \ddots & \sum_i\bra{t_{i+1} t_{i+1}}\rho_{\text{e-t}}\ket{t_{i+1} t_{i+1}} \cr 
\end{pmatrix}
\end{equation}

In this sense, the experimental results can be understood as averages over a larger state space in the energy-time domain. Henceforth we refer to the effective density matrix with time-delayed basis states $\ket{t}$ and $\ket{t+\tau}$: 

\begin{equation}
{{\rho}}_{\text{e-t}}'= \bordermatrix{
  & t_\text{A}t_\text{B}	& t_\text{A}(t_\text{B}+\tau)   & (t_\text{A}+\tau)t_\text{B}  &  (t_\text{A}+\tau) (t_\text{B}+\tau)   \cr
t_\text{A}t_\text{B} & \rho'_{0000} & \rho'_{0001} & \rho'_{0010} & \rho'_{0011} \cr
t_\text{A}(t_\text{B}+\tau)  & \ddots  & \rho'_{0101} & \rho'_{0110} & \rho'_{0111} \cr
(t_\text{A}+\tau)t_\text{B}  & \ddots & \ddots & \rho'_{1010} & \rho'_{1011} \cr
(t_\text{A}+\tau) (t_\text{B}+\tau) & c.c. &  \ddots & \ddots & \rho'_{1111} \cr
}
\end{equation}

In order to simplify the following discussion, we replace the time-shift operation, which would otherwise lead out of the two-dimensional energy-time subspace (i.e $\hat{\tau}\ket{t+\tau} \rightarrow  \ket{t+2\tau}$) with a NOT operation which transforms the new energy-time basis states $\ket{t}$ and $\ket{t+\tau}$ as

\begin{equation}
\begin{matrix}
\hat{\tau} \ket{t} &=& \ket{t+\tau} \\
\hat{\tau} \ket{t+\tau} &=& \ket{t} \\
\end{matrix}
\end{equation}

In doing so we neglect possible edge effects on timescales of the order of the pump coherence time. However, such effects would result in a decrease of coherence, such that they are without any consequence for our main objective of establishing lower bounds on entanglement. With this, the state of the effectively 4-dimensional system after traversing Alice's and Bob's calcite crystals can now  be written as: 

\begin{equation}
{\tilde{\rho}}_\text{pol,e-t}={\mathcal{T}}_\text{A} {\mathcal{T}}_\text{B} \left\{\ketbra{\Phi^+}{\Phi^+} \otimes {\rho}_\text{e-t}'\right\} {\mathcal{T}}_\text{A} {\mathcal{T}}_\text{B}
\end{equation}

Since the difference between adjacent time bins $\tau\sim \unit[2]{ps}$  was significantly shorter than the timing jitter of the detection system, the detection does not grant direct access to the exact emission time bins. Consequently we must perform a partial trace over the energy-time degree of freedom in order to obtain the reduced polarization density matrix (${\tilde{\rho}}_\text{pol,r}=\text{Tr}_\text{e-t}({\tilde{\rho}}_\text{pol,e-t})$) that can be accessed experimentally. The matrix can be written explicitly as:

\begin{equation}
{\tilde{\rho}}_\text{pol,r}
 =  \bordermatrix{
  & \text{H}_\text{A}\text{H}_\text{B}	& \text{H}_\text{A}\text{V}_\text{B}   & \text{V}_\text{A}\text{H}_\text{B}  &  \text{V}_\text{A}\text{V}_\text{B}    \cr
    \text{H}_\text{A}\text{H}_\text{B} & \frac{1}{2} & 0 & 0 & \frac{\text{Tr}(\hat{\tau}_\text{A} \hat{\tau}_\text{B}{\rho}_\text{e-t}')}{2} \cr
    \text{H}_\text{A}\text{V}_\text{B} & 0 & 0 & 0 & 0 \cr
    \text{V}_\text{A}\text{H}_\text{B} & 0 & 0 & 0 & 0\cr
    \text{V}_\text{A}\text{V}_\text{B} & \frac{\text{Tr}({\rho}_\text{e-t}' \hat{\tau}_\text{A} \hat{\tau}_\text{B})}{2}  & 0 & 0 &  \frac{1}{2} \cr
}
\end{equation}

where we have used $\text{Tr}({\rho})=1$, $\hat{\tau}^{-1}=\hat{\tau}$, and the invariance of the trace under cyclic permutation. As shown previously, the off-diagonal terms of the polarization density matrix, which determine the two-photon coherence, are now related to the off-diagonal terms of the energy-time density matrix:

\begin{equation}
\text{Tr}({\rho}'_\text{e-t} \hat{\tau}_\text{A} \hat{\tau}_\text{B}) = 2 \, \operatorname{Re}(\rho'_{0110} + \rho'_{1100})
\end{equation}

In other words, polarization coherence in the reduced polarization density matrix is a direct consequence of coherence in the energy-time domain. Note that the fringe visibility in the post-selection free Franson interferometers will also be limited by the fidelity of the polarization-entangled state compared to a maximally entangled state. Using this interpretation, the visibility of the polarization correlations impose a lower bound on the visibility in the energy-time domain. 

\begin{equation}\label{eq:viset}
 V_{\text{e-t}}^{\phi} =\text{max}_\phi\left[\langle{\sigma}^A_\phi \otimes {\sigma}^B_\text{x}\rangle\right]=   \text{max}_\phi\left[e^{i\phi}+e^{-i\phi}\right] \, \operatorname{Re}(\rho'_{1100} + \rho'_{0110} )
\end{equation}

\section*{Criteria for 2-dimensional entanglement based on experimental visibilities}
In the previous section we saw how individual visibility measurements in the polarization and energy-time subspace relate to certain elements of the respective subsystem density matrices. In the following we use these results to determine lower bounds on the concurrence $\mathcal{C}(  {\rho})$ and entanglement of formation $E_\text{oF}({\rho})$ in the polarization and energy-time-entangled subspaces. These values will then serve to establish a lower bound for the Bell-state fidelity $\mathcal{F}({\rho})$ of the entire system. In Ref. \cite{Huber:2011} easily computable lower bounds for the concurrence of mixed states that have an experimental implementation were derived. They are given by
\begin{equation}\label{marcusbound}
\mathcal{C}(\rho)\geq \frac{2\sqrt{2}}{\sqrt{d(d-1)}} \sum_{i,j>i}\left(|\bra{ii}\rho\ket{jj}|-\sqrt{\bra{ij}\rho\ket{ij}\bra{ji}\rho\ket{ji}}\right),
\end{equation}
where $d$ is the local dimension of $\rho$. It is possible to relax this lower bound by linearizing the above ineq. (\ref{marcusbound}). Since for all $a\in\mathbb{C}$ and for all $b,c\in\mathbb{R}$ it is true that $|a|\geq\text{Re}(a)$ and $\sqrt{bc}\geq\frac{1}{2}(b+c)$, it follows that 
\begin{equation}\label{exp1}
\mathcal{C}(\rho)\geq \frac{2\sqrt{2}}{\sqrt{d(d-1)}} \sum_{i,j>i}\left(\text{Re}\left(\bra{ii}\rho\ket{jj}\right)-\frac{1}{2}\left(\bra{ij}\rho\ket{ij}+\bra{ji}\rho\ket{ji}\right)\right).
\end{equation}
By defining the operator $W(d)$ that acts on a $d^2$-dimensional Hilbert space as
\begin{equation}
W(d)\coloneqq\sqrt{\frac{2}{d(d-1)}}\sum_{i,j>i}\left(2\ketbra{jj}{ii}-\ketbra{ij}{ij}-\ketbra{ji}{ji}\right),
\end{equation}
we can rewrite ineq. (\ref{exp1}) as
\begin{equation}\label{exp}
\mathcal{C}(\rho)\geq\text{Re}\left[\text{\normalfont Tr}(\rho W(d))\right].
\end{equation}

We define $\mathcal{C}_\text{lin}(\rho)$ as the right-hand side of the above ineq. (\ref{exp}), which is also a lower bound for the concurrence that can be measured experimentally. 

Another useful measure of entanglement is the entanglement of formation $E_\text{oF}(\rho)$, which represents the minimal number of maximally entangled bits (ebits) required to produce $\rho$ via an arbitrary local operations and classical communication (LOCC) procedure. It can be shown  \cite{Wootters:2001} that the entanglement of formation is lower bounded by the concurrence according to:

\begin{equation}\label{Eofdef}
E_\text{oF}(\rho)\geq-\log\left(1-\frac{\mathcal{C}(\rho)^2}{2}\right),
\end{equation} 

\subsection{Polarization entanglement criterion based on experimental visibility}
For the sake of brevity let us assume that the maximum correlation in Eq. \ref{rho_pol_Visibility_phi} is attained for $\phi=0$\footnote{Alternatively, one could incorporate the phase into the definition of the concurrence or the density matrix.}, so that:

\begin{equation}\label{rho_pol_Visibility_phi0}
 V_{\text{pol}}^{\phi}= 2\,[\operatorname{Re}(\rho_{0011}+\rho_{0110}) ]
\end{equation}

In the following, we show that the linear concurrence is lower bounded by the visibilities via:

\begin{equation}\label{eq:Clinpolbound}
 V_{\text{pol}}^{\phi} + V_{\text{pol}}^{\text{H}/\text{V}}-1 \leq \mathcal{C}_\text{lin}(\rho_\text{pol}) 
\end{equation} 

The experimental visibilities were $V^{\text{H}/\text{V}}_{\text{pol}}=99.33\pm0.015\%$ in the linear H/V measurement basis and $V^{\phi}_{\text{pol}}=98.5\pm 0.15\%$ in the coherent superposition basis, respectively. We can thus infer a lower bound $\mathcal{C}_\text{lin}(\rho_\text{pol}) \geq 0.9788 \pm 0.0015$  for the concurrence. Inserting into Eq. \ref{Eofdef} we see that this corresponds to a mininmum of $E_\text{oF}({\rho}_{\text{pol}})>0.94\pm0.004$ ebits of entanglement of formation.\\\\
\textbf{Proposition:} The concurrence in polarization space is lower bounded via Eq. \ref{eq:Clinpolbound}.\\\\
\textbf{Proof:} In the two-dimensional case the linearized concurrence (Eq. \ref{exp1})  reads:
\begin{equation}\label{Clinpol}
\mathcal{C}_\text{lin}(\rho_\text{pol})=2\, \text{Re}(\rho_{0011}) -\left( \rho_{1010}+\rho_{0101}\right) 
\end{equation}

Inserting \ref{Clinpol} on the r.h.s of \ref{eq:Clinpolbound} and using Eq. \ref{rho_pol_Visibility_HV} and Eq. \ref{rho_pol_Visibility_phi0} for the visibilities, we must show that:
\begin{equation}\label{xx}
 2\,\operatorname{Re}(\rho_{0011}) +2\,\operatorname{Re}(\rho_{0110})  + \rho_{0000} - \rho_{0101} - \rho_{1010} + \rho_{1111} -1 \leq 2\, \text{Re}(\rho_{0011}) -\left( \rho_{1010}+\rho_{0101}\right) 
\end{equation}

Re-ordering the terms, we obtain:

\begin{equation}
2\,\operatorname{Re}(\rho_{0110})  + \rho_{0000} + \rho_{1111} \leq 1
\end{equation}

Now, inserting 
\begin{equation}
\text{Tr}(\rho_{\text{pol}})=\rho_{0000} + \rho_{1111} + \rho_{1010}+\rho_{0101}=1
\end{equation}

on the r.h.s, the proof reduces to showing that:
\begin{equation}
\operatorname{Re}(\rho_{0110}) \leq \frac{1}{2}(\rho_{0101} + \rho_{1010})
\end{equation}

The l.h.s can be upper bounded using the Cauchy-Schwartz inequality: 
\begin{equation}\label{CS-ineq}
\operatorname{Re}(\rho_{0110}) \leq\Absolute{\rho_{0110}}\leq \sqrt{\rho_{1010}\rho_{0101}}
\end{equation}

which results in:
\begin{equation}\label{geometric-mean-ineq}
\sqrt{\rho_{1010} \rho_{0101}}\leq \frac{1}{2}(\rho_{1010}+\rho_{0101})
\end{equation}

which is true for any two positive numbers, thus concluding the proof.

\subsection{Energy-time entanglement criterion based on experimental visibility}
Since the adjacent time bins were too close to be resolved directly we did not measure the visibility in the computational basis. Hence we cannot use Eq. \ref{eq:Clinpolbound}. Note that a lower bound on the visibility in the computational basis could be obtained via the measurement of the coincidence-to-accidental ratio (CAR) in the experiment if we assume that path-length fluctuations due to atmospheric turbulence are  constant within the electronic coincidence window of 1ns. We chose to invoke a strictly weaker assumption: The matrix element $\rho'_{0110}$ is related to coherence of photons that were emitted from the crystal with a relative delay that exceeds the coherence time (i.e. accidental coincidences). So according to the definition of the coherence time, there is no phase relationship between these pairs and we can safely assume $\rho'_{0110}\approx0$. In other words, we assume that the free-space channel does not induce such coherence.  We believe that, while this assumption precludes a certification of entanglement that meets the requirements for quantum cryptography, it is physically meaningful and completely justified in our proof of concept experiment. Under this assumption Eq. \ref{eq:viset} reduces to:

\begin{equation}\label{eq:vis-et-2}
 V_{\text{e-t}}^{\phi} = 2\operatorname{Re}(\rho'_{1100} )
\end{equation}

In the following, we show that the linear concurrence is lower bounded via:
\begin{equation}\label{eq:clin-et-2}
2\, V_{\text{e-t}}^{\phi} -1 \leq \mathcal{C}_\text{lin}(\rho_\text{e-t})
\end{equation} 

Inserting the experimental Franson interference visibility of $V^{\phi}_{\text{e-t}}=95.6\pm 0.3\%$, we obtain $\mathcal{C}_\text{lin}(\rho_\text{e-t})\geq 0.912 \pm 0.006$ and $E_\text{oF}({\rho}_{\text{e-t}})>0.776\pm0.014$ ebits of entanglement of formation.\\\\
\textbf{Proposition:} The concurrence in the energy-time subspace is lower bounded via Eq. \ref{eq:clin-et-2}.\\\\
\textbf{Proof:} Inserting Eq. \ref{eq:vis-et-2} on the l.h.s and Eq. \ref{Clinpol} on the r.h.s of Eq. \ref{eq:clin-et-2}, we must show that:
\begin{equation}
4\operatorname{Re}(\rho'_{1100})-1 \leq 2\operatorname{Re}(\rho'_{1100})-(\rho'_{1010}+\rho'_{0101})
\end{equation}

 re-ordering we can write this as:

\begin{equation}
 2\operatorname{Re}(\rho'_{1100}) + (\rho'_{1010}+\rho'_{0101}) \leq 1\\
\end{equation}

Inserting 
\begin{equation}
\text{Tr}(\rho_{\text{e-t}})=\rho'_{0000} + \rho'_{1111} + \rho'_{1010}+\rho'_{0101}=1
\end{equation}

on the r.h.s, the proof reduces to showing that: 

\begin{equation}
\operatorname{Re}(\rho'_{1100}) \leq \frac{\rho'_{1111}+\rho'_{0000}}{2} 
\end{equation}

The l.h.s can be upper bounded using the Cauchy-Schwartz inequality: 
\begin{equation}\label{CS-ineq2}
\operatorname{Re}(\rho'_{1100}) \leq \Absolute{\rho_{1100}'}\leq \sqrt{\rho_{1111}'\rho'_{0000}}
\end{equation}

which results in:
\begin{equation}\label{geometric-mean-ineq2}
\sqrt{\rho'_{1111} \rho'_{0000}}\leq \frac{1}{2}(\rho'_{1111}+\rho'_{0000})
\end{equation}

which is true for any two positive numbers, thus concluding the proof.

\section*{Entanglement criterion for combined state space based on entanglement in subspaces}
In the previous section we used the results of individual visibility measurements in the polarization and energy-time subspace to determine a lower bound on the entanglement of formation $E_\text{oF}(\rho)$, concurrence $\mathcal{C}(\rho)$, and Bell-state fidelity $\mathcal{F}(\rho)$ in each subspace. In the following section we show that the results of these individual characterizations can be used to establish a lower bound on the entanglement of the entire hyperentangled system, and thus certify genuine high-dimensional entanglement. 

The concurrence of a pure bipartite state $\ket{\psi}$ acting on the finite-dimensional Hilbert space $\mathcal{H}_\text{A}\otimes\mathcal{H}_\text{B}$ is a measure of entanglement defined as \cite{Mintert:2005}
\begin{equation}
\mathcal{C}(\ket{\psi}) = \sqrt{2(1-\text{\normalfont Tr}(\rho_\text{A}^2))},
\end{equation}
where $\rho_\text{A}$ is the reduced state over the subspace $\mathcal{H}_\text{B}$. Its generalization for bipartite mixed states $\rho=\sum_ip_i\ketbra{\psi_i}{\psi_i}$ follows from the convex roof construction,
\begin{equation}
\mathcal{C}(\rho) = \inf_{\{p_i,\ket{\psi_i}\}} \sum_i p_i \mathcal{C}(\ket{\psi_i}),
\end{equation}
where the infimum is obtained over all possible decompositions of $\rho$. Given two bipartite mixed states, $\rho$ and $\sigma$, it follows from the definition of the concurrence the subadditivity relation
\begin{equation}
\mathcal{C}(\rho\otimes\sigma)\leq \mathcal{C}(\rho)+\mathcal{C}(\sigma).
\end{equation}

This quantity is in general very hard to compute, however, with Eq. \ref{exp} we have an easily computable lower bound $\mathcal{C}_\text{lin}(\rho)$ for the concurrence.

Let $\rho_\text{A}\in\mathcal{L}(\mathcal{H}^{d_\text{A}}\otimes\mathcal{H}^{d_\text{A}})$ and $\rho_\text{B}\in\mathcal{L}(\mathcal{H}^{d_\text{B}}\otimes\mathcal{H}^{d_\text{B}})$, be two unknown states whose values for $\mathcal{C}_\text{lin}(\rho_\text{A})$ and $\mathcal{C}_\text{lin}(\rho_\text{B})$ have been measured. Let $\rho_\text{AB}\in\mathcal{L}(\mathcal{H}^{d_\text{A}d_\text{B}}\otimes\mathcal{H}^{d_\text{A}d_\text{B}})$ be also an unknown state whose reduced states are $\rho_\text{A}=\text{\normalfont Tr}_\text{B}(\rho_\text{AB})$ and $\rho_\text{B}=\text{\normalfont Tr}_\text{A}(\rho_\text{AB})$, and whose concurrence $\mathcal{C}(\rho_\text{AB})$ one is interested in calculating. Because of the subadditivy character of the concurrence, it is not possible to simply add the known values of $\mathcal{C}_\text{lin}(\rho_\text{A})$ and $\mathcal{C}_\text{lin}(\rho_\text{B})$ in order to estimate $\mathcal{C}(\rho_\text{AB})$; it is necessary to calculate a lower bound for this quantity. We now show how to derive a useful lower bound for the concurrence $\mathcal{C}(\rho_\text{AB})$ given the experimentally accessible values $\mathcal{C}_\text{lin}(\rho_\text{A})$ and $\mathcal{C}_\text{lin}(\rho_\text{B})$.

First, notice that the reduced states $\rho_\text{A}$ and $\rho_\text{B}$ must satisfy
\begin{equation}\label{expab}
\mathcal{C}_\text{lin}(\rho_\text{A})=\text{Re}\left[\text{\normalfont Tr}(\rho_\text{A} W(d_\text{A}))\right]
\end{equation}
and
\begin{equation}\label{expa'b'}
\mathcal{C}_\text{lin}(\rho_\text{B})=\text{Re}\left[\text{\normalfont Tr}(\rho_\text{B} W(d_\text{B}))\right].
\end{equation}

Since $\mathcal{C}(\rho_\text{AB})\geq \mathcal{C}_\text{lin}(\rho_\text{AB})$, we calculate a lower bound for $\mathcal{C}(\rho_\text{AB})$ by minimizing $\mathcal{C}_\text{lin}(\rho_\text{AB})$ over all possible states $\rho_\text{AB}$ whose reduced states $\rho_\text{A}$ and $\rho_\text{B}$ satisfy conditions (\ref{expab}) and (\ref{expa'b'}). Namely,
\begin{equation}\label{minexp}
\mathcal{C}(\rho_\text{AB}) \geq \min_{\rho_\text{AB}} \text{Re}\left[\text{\normalfont Tr}(\rho_\text{AB} W(d_\text{A}d_\text{B}))\right].
\end{equation}

This minimization problem can now be solved by semi-definite programming (SDP). Defining $C_\text{lb}$ as the right-hand side of the above ineq. (\ref{minexp}), we write the following SDP:
\begin{align}
\begin{split}\label{sdp}
\text{given}				& \hspace{0.5cm} \mathcal{C}_\text{lin}(\rho_\text{A}),\mathcal{C}_\text{lin}(\rho_\text{B}) \\
C_\text{lb}=\min_{\rho_\text{AB}} 		& \hspace{0.5cm}  \text{Re}\left[\text{\normalfont Tr}(\rho_\text{AB} W(d_\text{A}d_\text{B}))\right] \\
\text{s.t.}					& \hspace{0.5cm} \rho_\text{AB}\geq0, \ \text{\normalfont Tr}(\rho_\text{AB})=1, \\
						& \hspace{0.5cm} \mathcal{C}_\text{lin}(\rho_\text{A}) =\text{Re}\left[\text{\normalfont Tr}(\text{\normalfont Tr}_\text{B}(\rho_\text{AB}) W(d_\text{A}))\right],\\
	 					& \hspace{0.5cm} \mathcal{C}_\text{lin}(\rho_\text{B}) =\text{Re}\left[\text{\normalfont Tr}(\text{\normalfont Tr}_\text{A}(\rho_\text{AB}) W(d_\text{B}))\right].							
\end{split}
\end{align}
The solution $C_\text{lb}$ of this SDP is a lower bound for $\mathcal{C}(\rho_\text{AB})$. 
Since the entanglement of formation $E_\text{oF}(\rho)$ is lower bounded by the concurrence according to
\begin{equation}
E_\text{oF}(\rho)\geq-\log\left(1-\frac{\mathcal{C}(\rho)^2}{2}\right),
\end{equation} 
the lower bound $C_\text{lb}$ for the concurrence $\mathcal{C}(\rho_\text{AB})$ allows us to also calculate a lower bound for the entanglement of formation $E_\text{oF}(\rho_\text{AB})$.

The fidelity to the $d$-dimensional maximally entangled state $\ket{\Phi^+_d}=\frac{1}{\sqrt{d}}\sum_i\ket{ii}$ can also be lower bounded from the subspace concurrences by altering the target function of the above SDP. Let $F_\text{lb}$ be
\begin{equation}
F_\text{lb} = \min_{\rho_\text{AB}} \text{\normalfont Tr}(\rho_\text{AB}\ketbra{\Phi^+_d}{\Phi^+_d}),
\end{equation}
where the minimum is taken over all states $\rho_\text{AB}$ that satisfy conditions (\ref{expab}) and (\ref{expa'b'}) for the subspace concurrences. Then, we can write the following SDP:
\begin{align}
\begin{split}\label{sdp2}
\text{given}				& \hspace{0.5cm} \mathcal{C}_\text{lin}(\rho_\text{A}),\mathcal{C}_\text{lin}(\rho_\text{B}) \\
F_\text{lb}=\min_{\rho_\text{AB}} 		& \hspace{0.5cm} \text{\normalfont Tr}(\rho_\text{AB}\ketbra{\Phi^+_d}{\Phi^+_d})\\
\text{s.t.}					& \hspace{0.5cm} \rho_\text{AB}\geq0, \ \text{\normalfont Tr}(\rho_\text{AB})=1, \\
						& \hspace{0.5cm} \mathcal{C}_\text{lin}(\rho_\text{A}) =\text{Re}\left[\text{\normalfont Tr}(\text{\normalfont Tr}_\text{B}(\rho_\text{AB}) W(d_\text{A}))\right],\\
	 					& \hspace{0.5cm} \mathcal{C}_\text{lin}(\rho_\text{B}) =\text{Re}\left[\text{\normalfont Tr}(\text{\normalfont Tr}_\text{A}(\rho_\text{AB}) W(d_\text{B}))\right].							
\end{split}
\end{align}
The solution $F_\text{lb}$ is a lower bound for the fidelity to the maximally entangled state $\mathcal{F}(\rho_\text{AB})\geq F_\text{lb}$.

For $\rho_\text{pol}$ and $\rho_\text{e-t}$ being entangled states in the polarization and time-energy degrees of freedom, respectively, we have showed how to calculate $\mathcal{C}_\text{lin}(\rho_\text{pol})$ and $\mathcal{C}_\text{lin}(\rho_\text{e-t})$ from the fringe visibilities $V_\text{pol}$ and $V_\text{e-t}$. For values of $\mathcal{C}_\text{lin}(\rho_\text{pol})=0.977$ and $\mathcal{C}_\text{lin}(\rho_\text{e-t})=0.906$, we obtain the lower bound $\mathcal{C}(\rho)\geq1.1299$ for the concurrence and $E_\text{oF}(\rho)\geq1.4671$ for the entanglement of formation of the hyperentangled state $\rho$. This bound is sufficient to guarantee $d=3$ entanglement \cite{Huber:2013}. For the same values of $\mathcal{C}_\text{lin}(\rho_\text{pol})$ and $\mathcal{C}_\text{lin}(\rho_\text{e-t})$ we obtain a lower bound for the fidelity to the maximally entangled state in $d=4$ of $\mathcal{F}(\rho)\geq0.9419$, which is enough to certify $d=4$ entanglement \cite{Fickler:2014}, an even higher dimensionality.

For these inputs values of $\mathcal{C}_\text{lin}(\rho_\text{pol})$ and $\mathcal{C}_\text{lin}(\rho_\text{e-t})$, and $d_\text{A}=d_\text{B}=2$, there exists a strictly feasible point $\rho^*$ that returns $\text{Re}\left[\text{Tr}(\rho^*W(4))\right]=1.14$ and $\text{Tr}(\rho^*\ketbra{\Phi^+_4}{\Phi^+_4})=0.95$ which guarantees that SDPs (\ref{sdp}) and (\ref{sdp2}) satisfy the condition of strong duality \cite{Boyd:2004}. This is proof that the optimal values $C_\text{lb}=1.1284$ and $F_\text{lb}=0.9410$ were indeed achieved by both primal and dual problems. Since these are finite values, it is guaranteed that the true minimum of both optimization problems was attained.

\section*{Supplementary References}

